\documentclass[12pt]{iopart}
\usepackage{iopams,graphicx,bm}  

\newcommand{\beq}{\begin{equation}}
\newcommand{\eeq}{\end{equation}}
\newcommand{\beqa}{\begin{eqnarray}}
\newcommand{\eeqa}{\end{eqnarray}}

\newcommand{\bfx}{{\bf x}}
\newcommand{\Mstar}{M^*}
\newcommand{\pairj}{j}
\newcommand{\psiup}{\psi_{\uparrow}}
\newcommand{\psidagup}{\psi^\dagger_{\uparrow}}
\newcommand{\psidown}{\psi_{\downarrow}}
\newcommand{\psidagdown}{\psi^\dagger_{\downarrow}}
\newcommand{\xvec}{{\bf x}}
\newcommand{\kf}{k_{\scriptscriptstyle\rm F}}
\newcommand{\grounds}{{\rm gs}}
\newcommand{\Gamint}{{\Gamma}_{\rm int}}
\newcommand{\galnab}{\stackrel{\leftrightarrow}{\nabla}}
\newcommand{\nab}{\overrightarrow{\nabla}}

\begin{document}

\title{Density Functional Theory: Methods and Problems}

\author{R J Furnstahl}

\address{Department of Physics, 
The Ohio State University, 
Columbus, OH\ \ 43210}

\ead{furnstahl.1@osu.edu}

\begin{abstract}
The application of density functional theory to nuclear structure 
is discussed, highlighting the current status of the effective
action approach using effective field theory, and outlining future
challenges. 
\end{abstract}


%\maketitle

\section{Introduction}

The similarities between 
self-consistent mean-field
approaches to nuclear structure, such as Skyrme Hartree--Fock
\cite{RINGSCHUCK,BENDER2003}, 
and Kohn-Sham density functional theory (DFT), which is widely applied
to Coulomb many-body problems \cite{DREIZLER90}, are frequently noted.
To go beyond the simple surface comparisons, however, there are
many questions to address about applying DFT methods to nuclear structure  
and potential problems or challenges one will confront in
trying to connect the two.
For example,
 \begin{itemize}
   \item How is Kohn-Sham DFT more than a mean-field calculation?  Where
   are the approximations?  How do we include long-range effects in a
   Skyrme-like formalism?
   \item What can you calculate in a DFT approach?  Excited states? 
   What about single-particle properties?
   \item How does pairing work in DFT?  Are higher-order contributions
   important?
   \item How do we deal with broken symmetries (translation, rotation,
   \ldots)?
   \item Can we connect quantitatively to the free NN interaction?  What
   about to chiral effective field theory (EFT) and/or low-momentum
   interactions?
 \end{itemize} 
In this talk, 
we discuss how to make the nuclear connection to DFT systematically,
by building a framework for addressing these questions.
The idea is to use effective actions of composite operators to develop
DFT \cite{FUKUDA}, 
with the Kohn-Sham orbitals
arising through the inversion method used to effect the
Legendre transformation to the effective action \cite{VALIEV,PUG02}.
The inversion method requires a hierarchy of approximations; here this is
provided by an effective field theory expansion, which uses power
counting to tell us what diagrams to include at each order \cite{HAMMER00}.
Ultimately, we seek nuclear energy functionals similar to conventional
``phenomenological'' approaches but model independent,
with error estimates, and derivable from microscopic EFT approaches.

\section{Effective Action Approach to EFT-Based Kohn-Sham DFT}

The dominant use of density functional theory (DFT) is in the 
description of interacting point electrons in
the static potential of atomic nuclei.
Applications include calculations of atoms, molecules, crystals, and
surfaces \cite{DREIZLER90}.
Discussions of DFT typically start with a proof by
contradiction of the existence of
the Hohenberg-Kohn (HK) energy functional 
\beq
  E_v[\rho] = F_{\rm HK}[\rho] + \int\!d^3x\, v({\bf x}) \rho({\bf x}) 
  \ ,
\eeq
which is minimized at the exact ground-state energy by the exact
ground-state density, and 
where $F_{\rm HK}[\rho]$ is \emph{universal}.  
The Kohn-Sham procedure further
posits a non-interacting system with the same
density as the fully interacting system \cite{DREIZLER90}.
This leads to constructing $E_v$ using single-particle 
orbitals in a 
local potential $v_{\rm KS}(\bfx)$,
\beq
  [-\bm{\nabla}^2/2m + v_{\rm KS}(\bfx)]\phi_i
  = {\varepsilon_i}\phi_i
   \ \Longrightarrow\ \rho(\bfx) = \sum_{i=1}^N |\phi_i(\bfx)|^2
   \ ,
   \label{eq:ksequation}
\eeq
which is determined self-consistently from $E_v[\rho]$.
The relative simplicity of solving Eq.~(\ref{eq:ksequation}) for large finite
systems is apparent while the
entire approach is a win if there are reasonable approximations to 
$F_{\rm HK}[\rho]$, such as the local density approximation (LDA).

We rely on a thermodynamic derivation of
DFT, which uses the effective action formalism \cite{NEGELE} 
to construct energy
density functionals \cite{FUKUDA,VALIEV}. 
The basic plan is to consider the partition function ${\cal Z}$ for
the (finite) system of interest in
the presence of external sources coupled to various quantities of
interest (such as the density).
We derive energy functionals of these quantities by Legendre
transformations with respect to the sources. 
These sources probe, in a
variational sense, configurations near the ground state.

To derive conventional
density functional theory, 
we consider an external source $J(x)$ coupled to the density operator 
$\widehat \rho(x) \equiv \psi^\dagger(x)\psi(x)$ in the partition
function
\beq
    {\cal Z}[J] = 
    e^{-W[J]} \sim {\rm Tr\,} 
      e^{-\beta (\widehat H + J\,\widehat \rho) }
    \sim \int\!{\cal D}[\psi^\dagger]{\cal D}[\psi]
    \,e^{-\int [{\cal L} + J\,\psi^\dagger\psi]} 
    \ ,
\eeq
for which we will construct a path integral representation
with Lagrangian ${\cal L}$ \cite{NEGELE}.
(Note: for convenience, we will take the inverse temperature
$\beta$ and the volume $V$ equal to unity in the sequel.)
The (time-dependent) density $\rho(x)$ in the presence of $J(x)$ is
\beq
  \rho(x) \equiv \langle \widehat \rho(x) \rangle_{J}
   = \frac{\delta W[J]}{\delta J(x)}
   \ ,
\eeq  
which we invert to find $J[\rho]$ and then Legendre transform from $J$ to
$\rho$:
\beq
   \Gamma[\rho] = - W[J] + \int\! J\, \rho
   \quad
   \mbox{with}
   \quad
   J(x) = \frac{\delta \Gamma[\rho]}{\delta \rho(x)}
   \longrightarrow 
   \left.\frac{\delta \Gamma[\rho]}{\delta \rho(x)}\right|_{\rho_{\rm gs}(\bfx)
   } =0
   \ .
\eeq 
For static $\rho(\bfx)$, $\Gamma[\rho]$ is proportional to 
the DFT energy functional $F_{\rm HK}$!  

We still need a way to carry out the inversion; for this we rely on the
inversion method of Fukuda et al. \cite{FUKUDA}.
The idea is to expand the relevant quantities in a hierarchy,
\beqa
   W[J,\lambda] &\!=\!& W_0[J] + \lambda W_1[J] + \lambda^2 W_2[J] + \cdots 
    \ , \\
   J[\rho,\lambda] &\!=\!& J_0[\rho] + \lambda J_1[\rho] + \lambda^2 J_2[\rho] 
      + \cdots \ , \\
   \Gamma[\rho,\lambda] &\!=\!& \Gamma_0[\rho] 
            + \lambda \Gamma_1[\rho] + \lambda^2 \Gamma_2[\rho] + \cdots
            \ , 
\eeqa
treating $\rho$ as order unity,
and match order by order in $\lambda$.  
Zeroth order is a noninteracting system with potential $J_0(x)$:
\beq
  \Gamma_0[\rho] = -W_0[J_0] + \int\!d^4x\, J_0(x)\rho(x) 
  \quad \Longrightarrow \quad \rho(x) = \frac{\delta W_0[J_0]}{\delta J_0(x)} 
  \ .   
  \label{eq:zeroth}
\eeq    
Because $\rho$ appears only at zeroth order, it is given
\emph{exactly} from the non-interacting system according 
to Eq.~(\ref{eq:zeroth}). 
This is the Kohn-Sham system with the exact density!
We diagonalize $W_0[J_0]$ by introducing Kohn-Sham orbitals as in
Eq.~(\ref{eq:ksequation}); it is
the sum of $\varepsilon_i$'s.
Finally, we find
$J_0$ for the ground state by completing 
a self-consistency loop:
\beq
  \fl
  {J_0} \rightarrow W_1 \rightarrow \Gamma_1 \rightarrow J_1
   \rightarrow W_2 \rightarrow \Gamma_2 \rightarrow \cdots
   \Longrightarrow
     {J_0(x) = 
     -\sum_{i>0} J_i(x) =
     \sum_{i>0} \frac{\delta\Gamma_{i}[\rho]}{\delta\rho(x)}} 
   \ .   
   \label{eq:loop}
\eeq
Note that even though solving for Kohn-Sham orbitals makes the approach
look like mean field, the approximation to the energy and density is 
\emph{only} in the truncation of Eq.~(\ref{eq:loop}) at some order.
Further, by using time-dependent sources we can generalize RPA to
time-dependent DFT, to calculate properties of collective excitations.

The hierarchy we have in mind is defined by EFT power counting.
For example, the EFT for a dilute Fermi system with short-range interactions 
is defined by a general interaction
as the sum of delta functions and their
derivatives. In momentum space,
\beq
  \langle {\bf k} | V_{\rm eft} | {\bf k'}\rangle
   = C_0 + \frac12 C_2 ({\bf k}^2 + {\bf k'}^2)
     + C'_2 {\bf k\cdot k'} + \cdots
     \ ,
\eeq
which corresponds to ${\cal L}_{\rm eft}$ with 
general contact interactions (including $3+$ bodies):
\beqa
  {\cal L}_{\rm eft} &=& 
       \psi^\dagger \Bigl[i\frac{\partial}{\partial t} 
               + \frac{\nab^{\,2}}{2M}\Bigr]
                 \psi - \frac{C_0}{2}(\psi^\dagger \psi)^2
            + \frac{C_2}{16}\bigl[ (\psi\psi)^\dagger 
                                  (\psi\!\galnab\!{}^2\psi)+\mbox{ h.c.} 
                             \bigr]   
  \nonumber \\[5pt]
   & & \null +
         \frac{C_2'}{8}(\psi\! \galnab\! \psi)^\dagger \cdot
              (\psi\!\galnab\!\psi)
   - \frac{D_0}{6}(\psi^\dagger \psi)^3 +  \ldots
   \ .
  \label{lag}
\eeqa
Dimensional analysis implies that
      $C_{2i} \sim \frac{\textstyle 4\pi}{\textstyle M } R^{2i+1}\,,
              D_{2i} \sim 
              \frac{\textstyle 4\pi}{\textstyle M} R^{2i+4} $,
where $R$ is the range of the interaction,
which gives our hierarchy (e.g., if $\kf R \ll 1$). 
This ${\cal L}_{\rm eft}$ will generate functionals like conventional Skyrme
functionals (but with a different power counting).

The conventional diagrammatic expansion of the propagator takes the
form
\begin{center}
  \includegraphics[angle=0.0,width=3.2in]{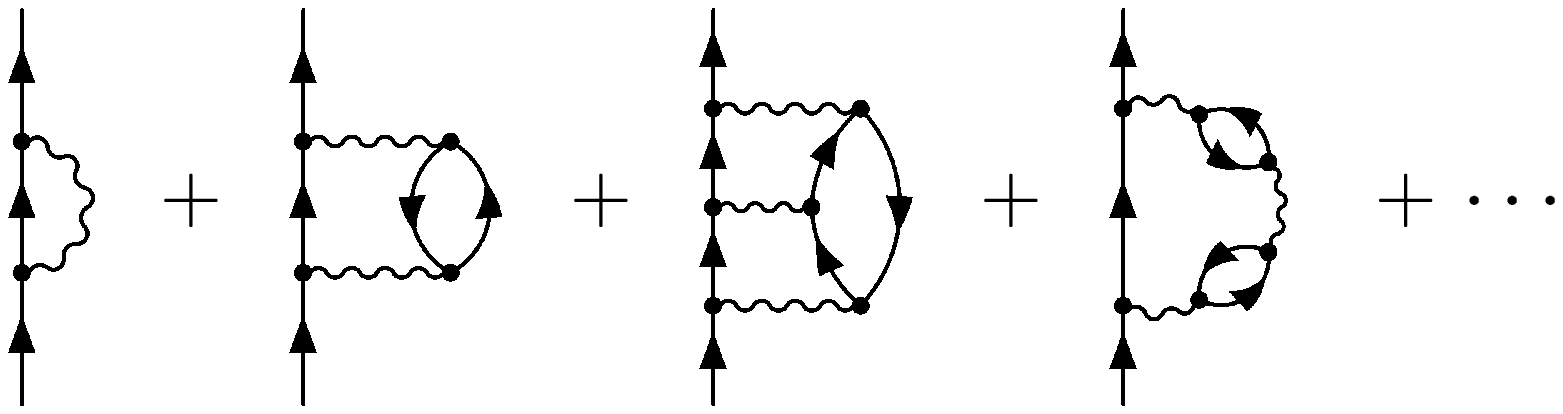}
     \ \raisebox{.41in}{$\Longrightarrow$}\
  \includegraphics[angle=0.0,width=2.5in]{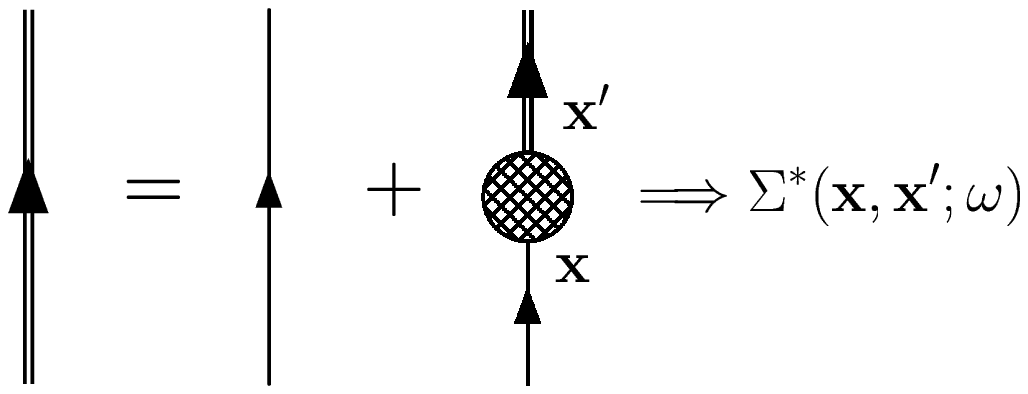}
\end{center}
with a
non-local, state-dependent 
$\Sigma^\ast({\bf x},{\bf x}';\omega)$.
In contrast, in Kohn-Sham DFT we have a \emph{local} $J_0({\bf x})$ to all
orders (i.e., not just at the Hartree level):
  \begin{center}
    \includegraphics[angle=0.0,width=5.8in]{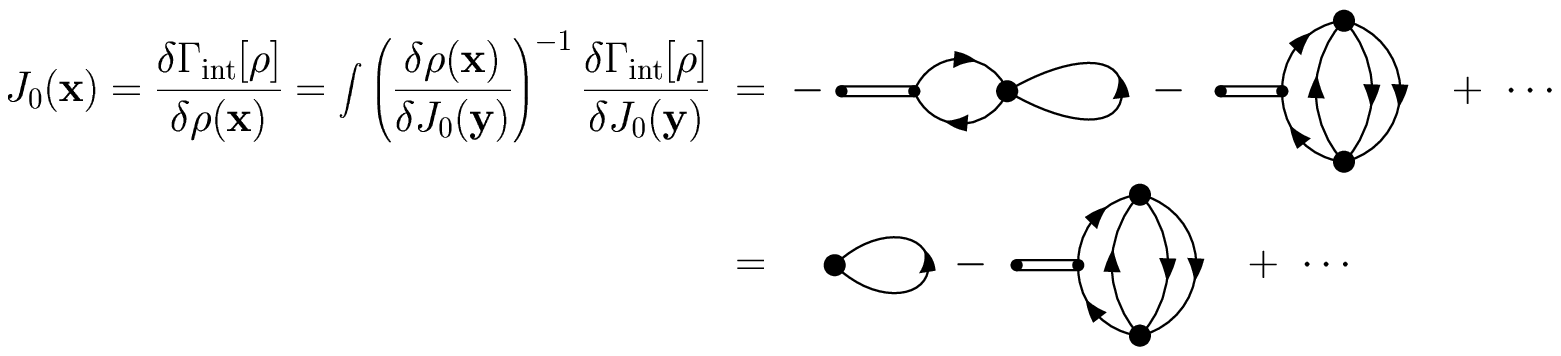}
  \end{center}
which introduces 
new Feynman rules for the ``inverse density-density correlator''
(double line) corresponding to $[\delta\rho(\bfx)/\delta J_0({\bf
y})]^{-1}$.
These inverse correlators appear in the energy,
\beq
   \hspace*{-.4in}
   \raisebox{-.25in}{\includegraphics[angle=0.0,width=5.2in]{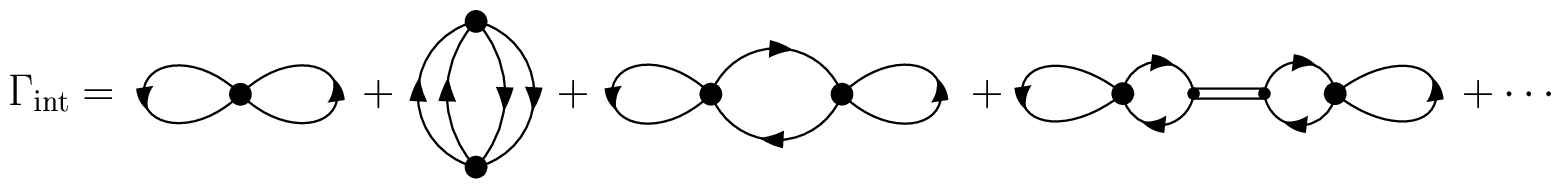}}
\eeq
but in this case 
the last two diagrams precisely cancel, in complete analogy to the
cancellation of anomalous diagrams according to Kohn, Luttinger, and
Ward \cite{NEGELE}.

We can show how the full Green's function $G$ is related to the Kohn-Sham
Green's function $G_{\rm ks}$ 
by adding a \emph{non-local} source $\xi(x,x')$ coupled to
$\psi(x)\psi^\dag(x')$ \cite{FURNSTAHL04b}: 
\begin{equation}
   \fl
    Z[J,\xi] = e^{-W[J,\xi]}
      = \int\! D\psi D\psi^\dag \ e^{-\int\! d^4x\ [{\cal L}\,+\,   J(x)
    \psi^\dag(x) \psi(x)\, {+\,
    \int\! d^4x'\, \psi(x)\xi(x,x')\psi^\dag(x')}]}
\end{equation}
The Green's function follows as a functional derivative of $W$ with
respect to $\xi$ at constant $J$, 
which is in turn equal to a functional derivative of
$\Gamma$ with respect to $\xi$ at constant $\rho$.
Since $\Gamma$ can be decomposed as $\Gamma_0$ from Eq.~(\ref{eq:zeroth})
plus everything else,
\begin{equation}
   \Gamma[\rho,\xi] = \Gamma_0[\rho,\xi] 
     + \Gamma_{\rm int}[\rho,\xi] \ ,
\end{equation}
and $G_{\rm ks}$ is the full Green's function for $\Gamma_0$,
we can derive a Dyson-like equation:
\begin{equation}
     G(x,x') = \left. \frac{\delta W}{\delta \xi}\right|_J
       = \left. \frac{\delta \Gamma}{\delta \xi}\right|_\rho
       = G_{\rm ks}(x,x') + G_{\rm ks}\Bigl[ 
       \frac{\delta\Gamma_{\rm int}}{\delta G_{\rm ks}}
       - \frac{\delta\Gamma_{\rm int}}{\delta \rho}
       \Bigr] G_{\rm ks}
       \ ,
\end{equation}
represented diagrammatically as:
\beq
  \raisebox{-.5in}{\includegraphics*[width=3.5in,angle=0]{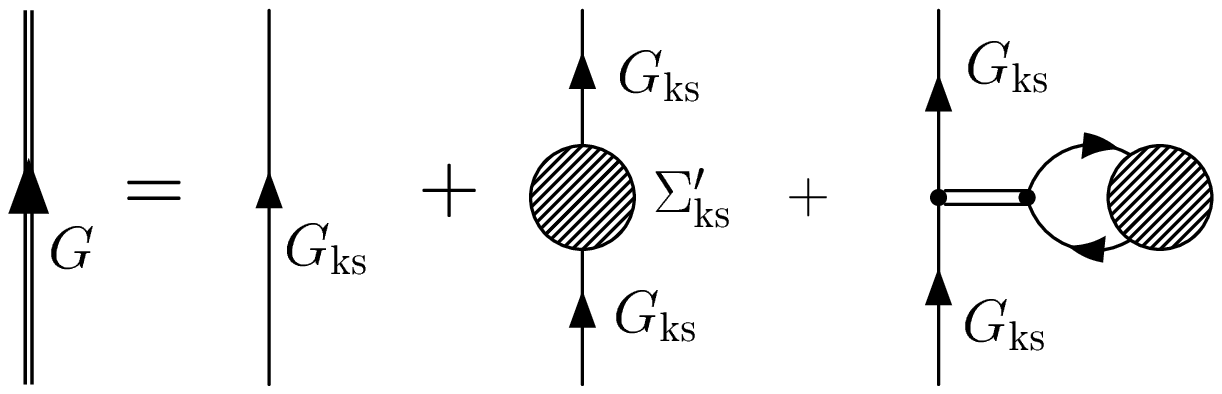}}
  \label{eq:fullG}
\eeq
We find a simple diagrammatic proof that the density from the exact
Green's function (by taking a trace) equals the density of the
non-interacting Kohn-Sham propagator:
\beq
  \raisebox{-.15in}{\includegraphics*[width=4.5in]{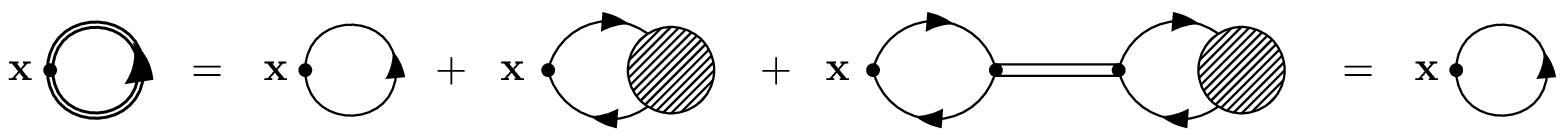}}
\eeq

\noindent
but other observables
quantities of interest, such as single-particle properties, will
generally differ and will need to be calculated using
Eq.~(\ref{eq:fullG}).

To study further how well single-particle properties are reproduced, 
we include the kinetic energy density
$\tau(x) \equiv \langle \bm{\nabla}\psi^\dagger\cdot\bm{\nabla}\psi \rangle$
in the DFT formulation for fermions in a trap.  
The meta-GGA DFT for the Coulomb problem
uses $\tau$ as an ingredient in the energy functionals, but applies
a semi-classical expansion in terms of the density \cite{DREIZLER90}.  
The result is a
Kohn-Sham equation with a local potential only.
In contrast, in Skyrme Hartree-Fock (SHF), $\rho$ and $\tau$ are treated as
independent ($N=Z$ nuclei here \cite{RINGSCHUCK}):
\beqa
  E_{\rm SHF}[\rho,\tau,{\bf J}] 
    &=& \int\!d^3x\,
    \biggl\{ {1\over 2M}\tau + {3\over 8} t_0 \rho^2
  + {1\over 16} t_3 \rho^{2+\alpha}
 + {1\over 16}(3 t_1 + 5 t_2) \rho \tau  \nonumber
  \\ & & \hspace*{-.1in}\null
  + {1\over 64} (9t_1 - 5t_2) (\bm{\nabla} \rho)^2  
  - {3\over 4} W_0 \rho \bm{\nabla}\cdot{\bf J}
  + {1\over 32}(t_1-t_2) {\bf J}^2 \biggr\}  
  \ ,
\eeqa
(${\bf J}$ is a spin-orbit density) and the orbital equations
now have effective mass $M^*(\bfx)$.

To generate the analogous functional in DFT/EFT, add to the Lagrangian 
$\eta({\bf x})\,\bm{\nabla}\psi^\dagger\bm{\nabla}\psi$ and
Legendre transform to an effective action of $\rho$ and $\tau$:
\beq
  \Gamma[\rho,\tau] = W[J,\eta]
   - \int\! J(x)\rho(x) - \int\! \eta(x)\tau(x) 
  \ .
\eeq
The inversion method results in two Kohn-Sham potentials,
\beq
     J_0({\bf x}) = \frac{\delta \Gamma_{\rm int}[\rho,\tau]}{\delta
     \rho({\bf x})} \quad \mbox{and} \quad 
     \eta_0({\bf x}) = \frac{\delta \Gamma_{\rm int}[\rho,\tau]}{\delta
     \tau({\bf x})} \ .     
\eeq
We diagonalize the quadratic part of the Lagrangian in $W_0$,
\beq
  \int\! d^4x\ \psi^\dagger \biggl[
  i\partial_t + \frac{\bm{\nabla}^{\,2}}{2M} -v(\xvec)+J_0(\xvec) 
  -\bm{\nabla}\cdot \eta_{0}(\xvec)\bm{\nabla}
  \biggr]\psi 
\eeq
and the Kohn-Sham equation becomes [with $v_{\rm KS} \equiv v(\xvec) -
J_0(\xvec)$]
\beq
%   \bigl[ -\frac{{\bm{\nabla}}^2}{2M}  +  v_{\rm KS}({\bf x})
%   \bigr]\, \psi_\alpha = \epsilon_\alpha \psi_\alpha
%   \ \Longrightarrow\
   \bigl[ 
   -\bm{\nabla}{\frac{1}{\Mstar({\bf{x}})}}\bm{\nabla}
     +  v_{\rm KS}({\bf x})
   \bigr]\, \psi_\alpha = \epsilon_\alpha \psi_\alpha
\eeq         
with an effective mass
$1/2\Mstar({\bf x}) \equiv 1/2M - \eta_0({\bf x})$,
just like in Skyrme HF \cite{FURNSTAHL04b}. 

To highlight the differences between an energy functional of $\rho$
alone and one of $\rho$ and $\tau$, we focus on the Hartree-Fock
contributions from vertices with derivatives:
\beq
  {\cal L}_{\rm eft} = \ldots
       + \frac{C_2}{16}\bigl[ (\psi\psi)^\dagger 
                                  (\psi\!\galnab\!{}^2\psi)+\mbox{ h.c.} 
                             \bigr]   
    +
         \frac{C_2'}{8}(\psi\! \galnab\! \psi)^\dagger \cdot
              (\psi\!\galnab\!\psi) + \ldots
   \ .
\eeq
The corresponding part of the 
energy density with spin degeneracy $\nu=2$ for Kohn-Sham LDA is
\cite{PUG02}
 \beq
   {\cal E}_{\rm int}[\rho] = \ldots + 
     \frac{C_2}{8}
     \Bigl[
       \frac35\left(\frac{6\pi^2}{\nu}\right)^{2/3}\rho^{8/3}
     \Bigr]
    +
     \frac{3C_2'}{8}
     \Bigl[
       \frac35\left(\frac{6\pi^2}{\nu}\right)^{2/3}\rho^{8/3}
     \Bigr]
     + \ldots
 \eeq
and with $\tau$ is \cite{FURNSTAHL04}
\beq
  {\cal E}_{\rm int}[\rho,\tau] = \ldots +
   \frac{C_2}{8}
   \bigl[
     \rho\tau + \frac34(\bm{\nabla}\rho)^2
   \bigr]
  +
   \frac{3C_2'}{8}
   \bigl[
     \rho\tau - \frac14(\bm{\nabla}\rho)^2
   \bigr]
  + \ldots
  \ .
  \label{eq:Erhotau}
\eeq
The $\rho\tau$ functional with $\nu=4$
takes the same form as the Skyrme functional with
$C_i \rightarrow t_i$.  (To get the spin-orbit part, one should
couple a source to
 the spin-orbit density.)

\begin{figure}[t]
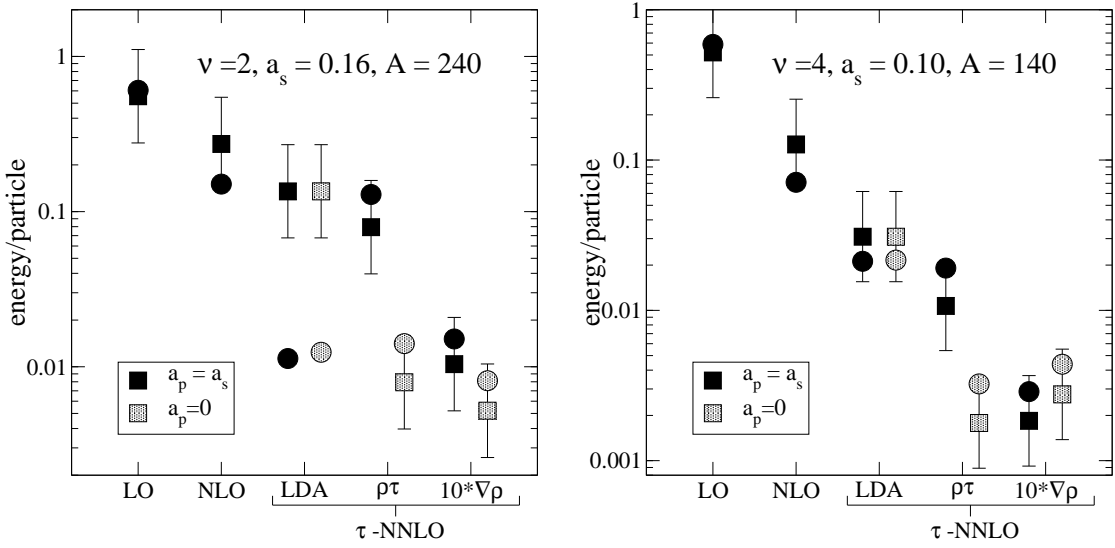

\centerline{\includegraphics*[width=2.8in,angle=0]{error_plot_tau_breakup240}
            \hspace*{.1in}
            \includegraphics*[width=2.8in,angle=0]{error_plot_tau_breakup140}}
\vspace*{-.1in}
\caption{Contributions to the energy for two systems of trapped
fermions \cite{FURNSTAHL04}.}
\label{fig:errorplots}
\end{figure}        

First, we note that power counting estimates work for all terms in
the $\rho\tau$ energy functional, as shown for two examples in
Fig.~\ref{fig:errorplots} [the ``$\rho\tau$'' and
``$\nabla\rho$'' contributions are from Eq.~(\ref{eq:Erhotau})].
The squares with error bars are estimated contributions to the energy
per particle from terms in the energy functional at each order.  The
error bars indicate a ``natural'' range of coefficients between 1/2 and
2.  The circles are the actual values.  The hierarchy as well as the 
accuracy of the predictions is evident \cite{FURNSTAHL04}. 

\begin{figure}[b]
\centerline{%      
       \begin{tabular}{ccc}
         {$a_p = a_s$} & $E/A$ & $\sqrt{\langle r^2\rangle}$ \\ \hline
         $\rho$ & 7.66 & 2.87 \\
         $\rho\tau$ & 7.65 & 2.87 \\ \hline \\[5pt]
         {$a_p = 2 a_s$} & $E/A$ & $\sqrt{\langle r^2\rangle}$ \\ \hline
         $\rho$ & 8.33 & 3.10 \\
         $\rho\tau$ & 8.30 & 3.09 \\ \hline
       \end{tabular}
       \hspace*{.2in}
       \raisebox{-1.3in}{%
       \includegraphics*[width=3.7in,angle=0]{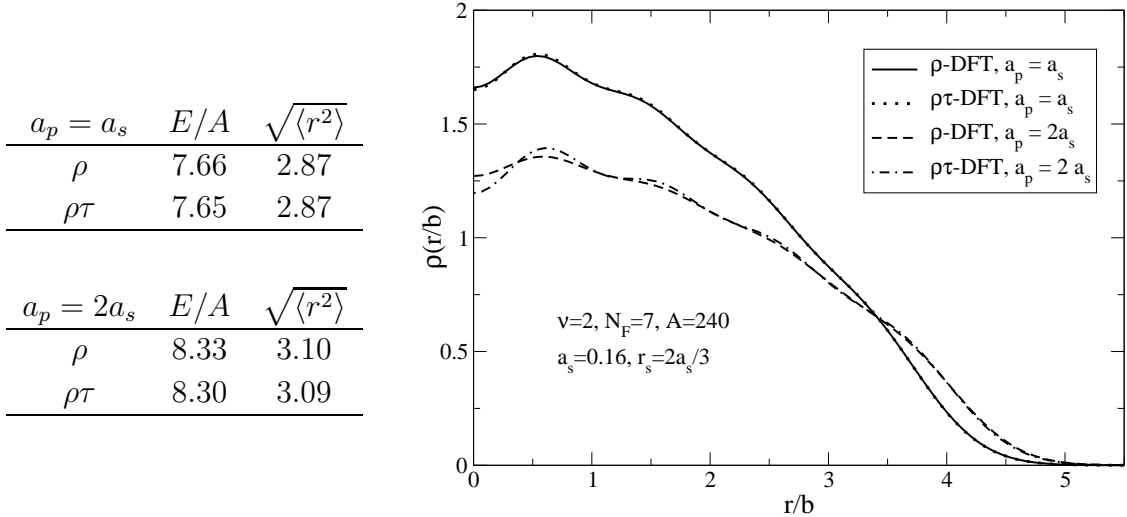}}}
%\vspace*{-.1in}
\caption{Results for the $\rho$ and $\rho\tau$ functionals
for  representative cases \cite{FURNSTAHL04}.}
\label{fig:comparison}
\end{figure}        

\begin{figure}[t]
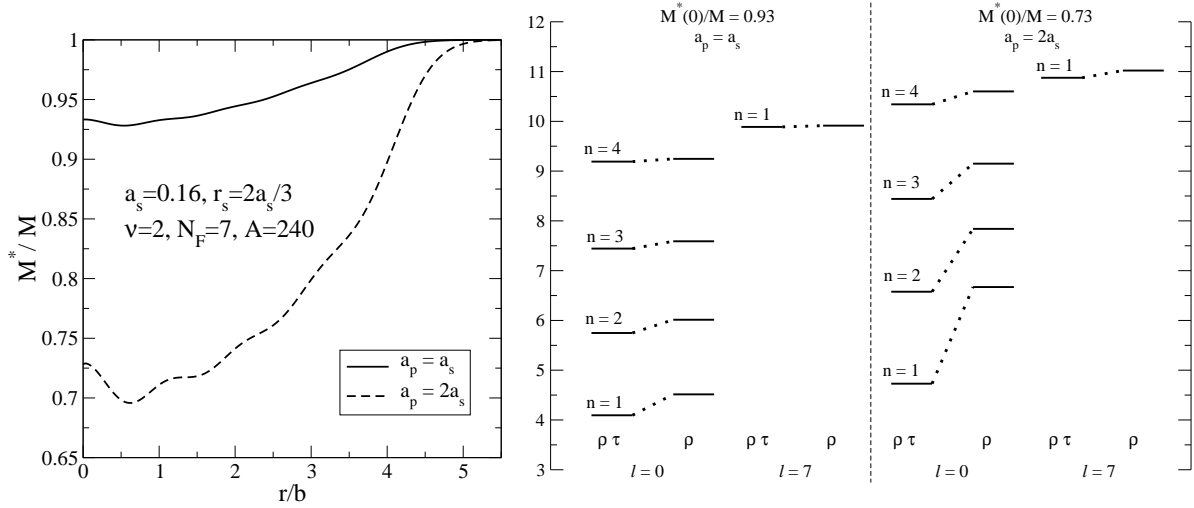

\centerline{\includegraphics*[width=2.6in,angle=0]{EffectiveMass240n}
            \hspace*{.0in}
            \raisebox{.1in}{%
            \includegraphics*[width=3.5in,angle=0]{EnergySpectrum240b}}}
\vspace*{-.1in}
\caption{a) Effective mass and b) single-particle spectra for 
representative cases \cite{FURNSTAHL04}.}
\label{fig:singleparticle}
\end{figure}

We also note from these figures that the contributions that are new to
the $\rho\tau$ functional are very small.  Thus, we are not
surprised to find that predicted bulk observables are very similar (see
Fig.~\ref{fig:comparison}).  The total binding energy and density
are what Kohn-Sham is supposed to get right.
What about the single-particle spectrum?  The effective mass is unity
in the $\rho$ case and, as seen in Fig.~\ref{fig:singleparticle}(a),
is reduced significantly in the $\rho\tau$ case (the central values
roughly cover the range in typical Skyrme interactions).  The different
effective mass is reflected in the single-particle spectra in
Fig.~\ref{fig:singleparticle}(b), which can be understood qualitatively
from
the spectra calculated in a uniform system,
\beq
  \varepsilon_{\bf k}^{\rho} - \varepsilon_{\bf
         k}^{\rho\tau}
          = \frac{\pi}{\nu}[(\nu-1) a_s^2 r_s + 2(\nu+1) a_p^3]\, 
          \frac{k_{F}^2 - {\bf k}^2}{2M}\rho
       \ .
\eeq
Even though the density and energy are the same, the spectra are
not.
However, the $\rho\tau$ result is closer to that of
the full Green's function, as can be shown from Eq.~(\ref{eq:fullG})
\cite{FURNSTAHL04}.

An important ingredient of energy functionals for nuclei
is pairing, which requires another generalization of Kohn-Sham DFT.
To do so,
we introduce an external source $j$ coupled to the pair density, to break
the $U(1)$ phase symmetry associated with fermion number.  
The generating functional now becomes
\beq
    Z[J,\/\pairj] = e^{-W[J,\/\pairj]}
      = \int\! D(\psi^\dag\psi) 
      \ e^{-\int\! d^4x\ [{\cal L}\,+\,   J(x)
    \psi_\alpha^\dag \psi_\alpha
    \,+\, \pairj(x)(\psidagup\psidagdown + \psidown\psiup)]}
    \ ,
\eeq
and the fermion and pair
densities $\rho,\phi$ are found by functional derivatives with respect
to $J,\pairj$:
\beqa
  \rho(x) &\equiv& \langle \psi^\dag(x)\psi (x)\rangle_{J,\pairj}
  = \left.\frac{\delta W[J,\pairj]}{\delta J(x)}\right|_{\pairj}
    \ ,
  \\ 
      \phi(x) &\equiv& \langle 
        \psidagup(x)\psidagdown(x)+ \psidown(x)\psiup(x)
      \rangle_{J,\pairj}
      = \left.\frac{\delta W[J,\pairj]}{\delta \pairj(x)}\right|_{J}
    \ .
\eeqa
The effective action $\Gamma[\rho,\phi]$
follows yet again by functional Legendre transformation: 
\beq
    \Gamma[\rho,\phi]  = 
      W[J,\pairj] - \int\! d^4x\, J(x)\rho(x)
      - \int\! d^4x\, \pairj(x)\phi(x)
      \ ,
\eeq
and is proportional to the ground-state energy functional.
The sources in turn are given by functionals derivatives with respect
to $\rho$ and $\phi$, which are therefore determined in the ground state
(where the sources are zero) by stationarity:
\beq
   \left.
     \frac{\delta \Gamma[\rho,\phi]}{\delta \rho({\bf x})}
    \right|_{{\rho=\rho_\grounds,\phi=\phi_\grounds}}
    =
   \left.
     \frac{\delta \Gamma[\rho,\phi]}{\delta \phi({\bf x})}
    \right|_{{\rho=\rho_\grounds,\phi=\phi_\grounds}}
    = 0
  \ .
\eeq
This is Hohenberg-Kohn DFT extended to pairing!

Inversion is again carried out by order-by-order matching 
in an EFT expansion.
Zeroth order is the Kohn-Sham system with potentials 
$J_0({\bf x})$ and $\pairj_0({\bf x})$, which yields the
\emph{exact} densities $\rho({\bf x})$ and $\phi({\bf x})$.
We introduce single-particle orbitals and solve
\beq
 \left(
   \begin{array}{cc}
    h_0(\xvec) - \mu_0 & \pairj_0(\xvec) \\
    \pairj_0(\xvec)         & -h_0(\xvec) + \mu_0 
   \end{array}
 \right)
 \left(
   \begin{array}{c}
   u_i(\xvec) \\ v_i(\xvec)
   \end{array}
 \right)
 = E_i
 \left(
   \begin{array}{c}
   u_i(\xvec) \\ v_i(\xvec)
   \end{array}
 \right)
\eeq
where $u_i,v_i$ satisfy conventional orthonormality relations 
\cite{RINGSCHUCK,BENDER2003} and
\beq
   h_0 (\xvec) \equiv -\frac{\bm{\nabla}^2}{2M} + v(\xvec)
      -J_0(\xvec)
      \ .
\eeq
Thus it looks like conventional Hartree-Fock-Bogoliubov (HFB)!
    
The diagrammatic expansion is the same as without pairing, 
but now uses Nambu-Gorkov matrix
Green's functions.  The Kohn-Sham self-consistency procedure is the same
as in Skyrme or RMF with pairing \cite{RINGSCHUCK,BENDER2003}.
In terms of the orbitals, the fermion density is given by
\beq
  \rho(\xvec) =
  2\sum_i\, |v_i(\xvec)|^2 
  \ ,
\eeq
and the (bare, unrenormalized) pair density is 
\beq
  \phi_{\rm bare}(\xvec) =
  \sum_i\, [ u_i^*(\xvec) v_i(\xvec) + u_i(\xvec) v_i^*(\xvec) ]
  = 2\sum_i u_i(\xvec)v_i(\xvec)
  \ .
\eeq
The chemical potential $\mu_0$ is fixed by $\int\!\rho(\xvec) = A$
and diagrams for
$\Gamma[\rho,\phi] = -E[\rho,\phi]$ 
yields Kohn-Sham potentials
\beq
    J_0(\xvec)\Bigr|_{\rho=\rho_{\rm gs}} = 
      \left.\frac{\delta \Gamint[\rho,\phi]}{\delta \rho(\xvec)}
    \right|_{\rho=\rho_{\rm gs}}
  \  \mbox{and} \quad
    \pairj_0(\xvec)\Bigr|_{\phi=\phi_{\rm gs}} = 
      \left.\frac{\delta \Gamint[\rho,\phi]}{\delta \phi(\xvec)}
    \right|_{\phi=\phi_{\rm gs}}
    \ .
\eeq

The pair density is divergent and requires renormalization.
This problem is solved for a uniform system in Ref.~\cite{FURNSTAHL05} 
by introducing
a counterterm proportional to $j^2$.
In this uniform limit, $\phi$ is ultimately defined with a subtraction:
\beq
     \phi = \int^{k_c}\! \frac{d^3k}{(2\pi)^3} \, j_0 
        \left(
          \frac{1}{\sqrt{(\epsilon_k^0-\mu_0)^2 + j_0^2}}
          - \frac{1}{\epsilon_k^0}
        \right)
    \stackrel{k_c\rightarrow \infty}{\longrightarrow} \mbox{finite}
    \ ,
\eeq
which can be applied in a local density approximation (i.e.,
Thomas-Fermi), 
\beq
  \fl 
  \phi(\xvec) = 2\sum_i^{E_c} u_i(\xvec)v_i(\xvec)
     - j_0(\xvec) \frac{M k_c(\xvec)}{2\pi^2}
  \quad \mbox{with} \quad
  E_c = \frac{k_c^2(\xvec)}{2M} + v_{\rm KS}(\xvec) - \mu
  \ .
\eeq
However, convergence is very slow as the energy cutoff $E_c$ is increased,
which is a practical problem for implementation in nuclei.
Bulgac and Yu have shown how one can make different subtractions and
greatly improve the convergence \cite{BULGAC}.  
For example, defining $\phi$ instead
through
\beq
  \phi = \int^{k_c}\! \frac{d^3k}{(2\pi)^3} \, j_0 
     \left(
       \frac{1}{\sqrt{(\epsilon_k^0-\mu_0)^2 + j_0^2}}
       - \frac{{\cal P}}{\epsilon_k^0-\mu_0}
     \right)
 \stackrel{k_c\rightarrow \infty}{\longrightarrow} \mbox{finite}
\eeq
leads to much faster convergence, as seen here for the uniform system:

\centerline{\includegraphics*[angle=0.0,width=3.1in]{convergence_plot}}

\noindent
Bulgac has found even faster convergence with other subtractions.

A question to be investigated is the role of higher-order contributions.
In the weak-coupling limit, the pairing gap in a dilute Fermi gas
with two spin states
is reduced approximately fifty percent by induced interactions.
The DFT/EFT framework is perfectly suited to explore the ramifications
for finite nuclei, with a consistent treatment of interactions
in the particle-particle 
and particle-hole channels.

\section{Problems and Challenges}

There are many problems and challenges 
facing the DFT/EFT program for nuclei.  
Here we give a personal laundry list of
what we believe are priority issues.

\begin{itemize}
  \item {\bf Including long-range effects.}  Two classes of
  long-range phenomena are particularly relevant.  On the left, we have  
  long-range forces, such as pion exchange.
    
  \includegraphics[width=2.8in,angle=0]{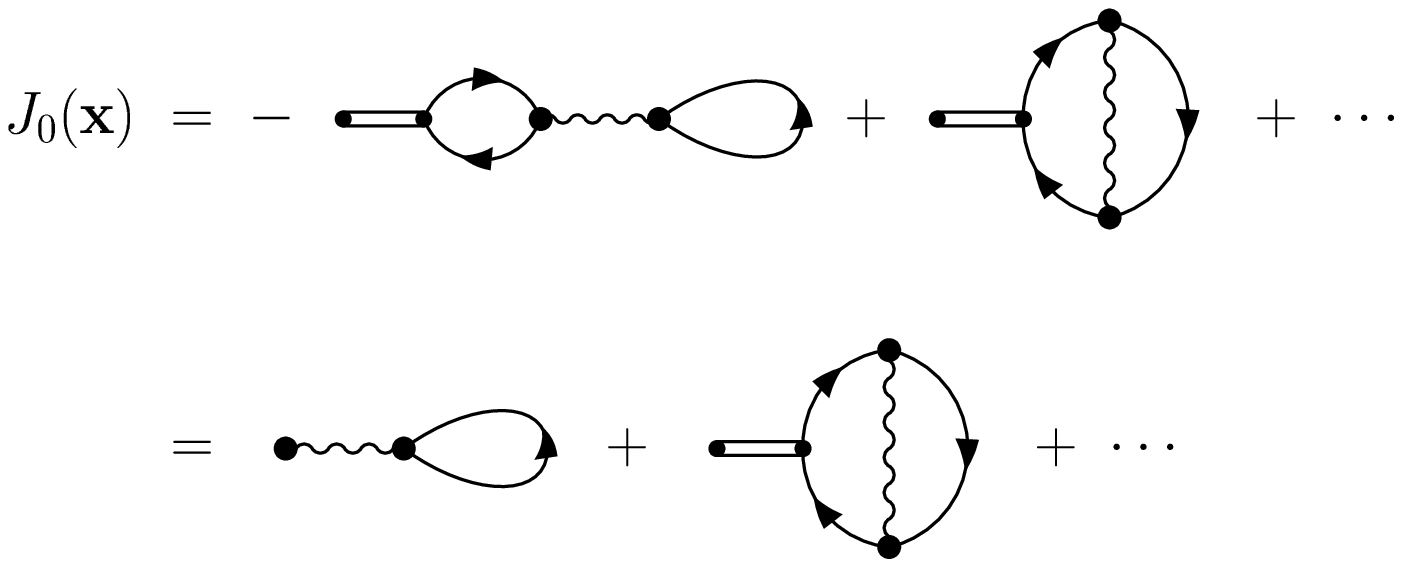}
  \hspace*{.1in}
  \raisebox{.3in}{\includegraphics[width=2.8in,angle=0]{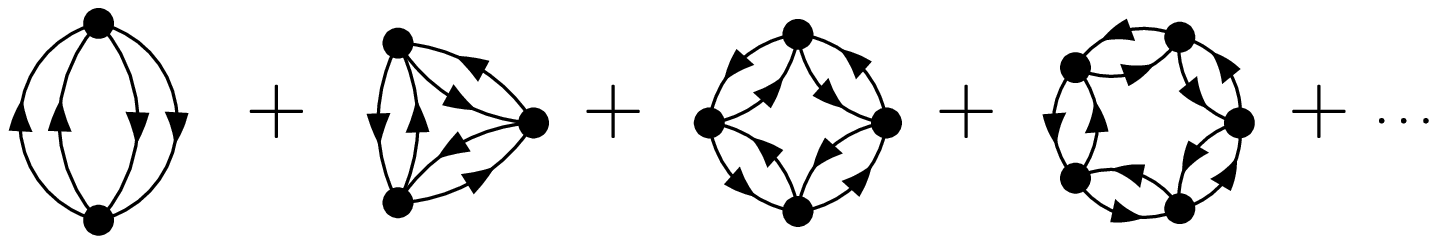}}
  
  On the right are diagrams leading
  to non-localities from near-on-shell
  particle-hole excitations.  We will need appropriate expansions or
  resummations for both.  
  
  \item {\bf  Developing gradient expansions.}  
  We need to consider semiclassical
  expansions used in Coulomb DFT and to revisit the density matrix
  expansion.  Furthermore, there are gradient expansion techniques for
  one-loop effective actions, which might be adapted.  The first
  targets are
  the beachball diagram for short-range forces and the Fock diagram for
  pion exchange.
  
  \item {\bf Restoring broken symmetries.}
  In Coulomb applications, a fixed external potential (nuclear Coulomb
  field) means that translational and rotational invariance are not
  issues, so they have not been studied via DFT.  This is also true of
  particle-number violation in HFB.  In the effective action framework,
  these correspond to problems from ``zero modes,'' which can be dealt
  with using a Fadeev-Popov
  approach.  Work in progress seeks an energy
  functional for the intrinsic density, using a one-dimensional toy
  model as a theoretical laboratory.

  \item {\bf Developing auxiliary field Kohn-Sham theory.}
  Auxiliary fields are like non-dynamical meson fields, which can be
  incorporated in the formalism using an appropriate saddlepoint
  evaluation.  It may be profitable to revisit the large $N$ expansion
  in this context.  For pairing, the challenge is to separate
  particle-hole and particle-particle channels for short-range
  interactions.
  
  \item {\bf Studying covariant DFT.}
  Relativistic mean-field models also need to be cast into DFT form.
  We are exploring a controlled laboratory for finite systems using
  short-range interactions only.
  Special renormalizations allow for great simplifications.  
  The connection to three-body forces
   in the nonrelativistic limit is being pursued, 
  as well as covariant pairing and
  time-dependent Kohn-Sham theory.   
\end{itemize}
\noindent
Work in all of these areas is in progress.

Perhaps the ultimate challenges for the DFT program for nuclear structure is to
connect quantitatively
to microscopic approaches.  Past attempts to derive Skyrme
parameters from NN interactions have resulted in qualitative or
semi-quantitative agreement, but have always fallen short of useful
quantitative predictions.
In contrast, Kohn-Sham density functional theory for Coulomb systems
is ``ab initio'' in the sense that the LDA part of the energy functional
comes directly from numerical calculations (with a form based on analytic
limits) of a uniform interacting electron gas.  

The successes of Coulomb DFT are often attributed to the fact that
Hartree-Fock gives the dominate contribution, so that correlations are
small corrections (even though the precision required --- chemical
accuracy --- is high!).  In turn, it has been said that the same level
of success is unlikely for nuclear systems since for 
\emph{conventional} NN
interactions, the correlations are larger than Hartree-Fock.
However, these conventional potentials are not the only choice for
many-body interactions.

If we start with a chiral effective potential and then run a momentum
cutoff down, we end up with a much softer potential.
In the medium, as the
cutoff is run toward the Fermi surface,      
the phase space for particle-particle scattering intermediate states in
the medium becomes severely constrained.
The apparent result is a more perturbative many-body system
(at least in the particle-particle
channel), so Hartree-Fock (with both two- and three-body low-momentum
interactions) 
is a useful starting point \cite{SCHWENK}.
Deriving constants for Skyrme-like energy functionals from this approach
is an important problem for the near future.

\ack

I gratefully acknowledge my collaborators A.~Bhattacharyya, S.~Bogner,
J.~Engel,
H.-W.~Hammer, S.~Puglia,
A.~Schwenk, and B.~Serot.
This work is supported by the NSF under Grants No.~PHY--0098645 
and No.~PHY--0354916.

\end{document}